\documentclass{article}

\usepackage{amssymb,amsmath}
\usepackage{graphicx,subfigure}

\newcommand{\Real}{\mathbb{R}}

\newcommand{\Y}{\mathbf{Y}}
\newcommand{\Q}{\mathbf{Q}}
\newcommand{\X}{\mathbf{X}}
\newcommand{\Z}{\mathbf{Z}}
\newcommand{\W}{\mathbf{W}}
\newcommand{\F}{\mathbf{F}}
\newcommand{\A}{\mathbf{A}}
\newcommand{\B}{\mathbf{B}}
\newcommand{\C}{\mathbf{C}}
\newcommand{\G}{\mathbf{G}}
\newcommand{\R}{\mathbf{R}}
\newcommand{\M}{\mathbf{M}}
\newcommand{\overlap}{\mathbf{S}}
\newcommand{\V}{\mathbf{V}}
\newcommand{\U}{\mathbf{U}}
\newcommand{\id}{\mathbf{I}}

\newcommand{\zero}{\mathbf{0}}
\newcommand{\position}{\mathbf{r}}
\newcommand{\Position}{\mathbf{R}}
\newcommand{\direction}{\mathbf{P}}

\newcommand{\extpotential}{\mathbf{v}}
\newcommand{\intpotential}{\mathbf{P}}
\newcommand{\laplace}{\mathbf{L}}
\newcommand{\density}{\mathbf{n}}

\newcommand{\step}{\tau}
\newcommand{\householder}{\mathbf{H}}
\newcommand{\transport}{\mathbf{T}}

\newcommand{\tangentspace}[1]{\mathcal{T}_{#1}\mathcal{M}}

\newcommand{\manifold}{\mathcal{M}}

\title{Direct minimization of electronic structure calculations with Householder reflections}

\author{K. Baarman\footnote{Department of Mathematics and Systems Analysis, Aalto University School of Science, Espoo, Finland, e-mail: \texttt{kurt.baarman@aalto.fi}}, T. Eirola\(^*\), and V. Havu\footnote{Department of Applied Physics, Aalto University School of Science, Espoo, Finland}}

\date{\today}
\begin{document}
\maketitle

\begin{abstract}
We consider a minimization scheme based on the Householder transport operator for the Grassman manifold, where a point on the manifold is represented by an \(m\times n\) matrix with orthonormal columns.
In particular, we consider the case where \(m \gg n\) and present a method with asymptotic complexity \(mn^2\).
To avoid explicit parametrization of the manifold we use Householder transforms to move on the manifold, and present a formulation for simultaneous Householder reflections for  \(\overlap\)-orthonormal columns.
We compare a quasi-Newton and nonlinear conjugate gradient implementation adapted to the manifold with a projected nonlinear conjugate gradient method, and demonstrate that the convergence rate is significantly improved if the manifold is taken into account when designing the optimization procedure.
%\keywords{Quasi-Newton method \and minimization \and electronic structure \and Householder transform}
\end{abstract}

%\begin{AMS}
%49M15, 65K10, 65Z05
%\end{AMS}

\section{Introduction}
We consider the optimization problem
\begin{equation}\label{eq:minimization}
\min_{\X^T\X=\id} f(\X),
\end{equation}
that is, we attempt to minimize the real valued function \(f\) of \(\X\in\Real^{m\times n}\) where \(m \gg n\), subject to the constraint \(\X^T\X = \id\), and with the computable derivative \(df(\X)\). The constraint on \(\X\) ensures that \(f\) has a minimum, but this minimum is not necessarily unique.
The method we present requires that \(m \ge 2n\). Extending the method to cover cases where just \(m > n\) is, however, possible.

A special property we assume of \(f\) is the homogeneity condition: \(f(\X) = f(\X\Q)\), where \(\Q\) is any \(n\times n\) orthonormal matrix. This property means that \(f\) only depends on the span of the columns of \(\X\).
The set of subspaces that are spanned by the columns of orthonormal \(m\times n\) matrices is called the Grassman manifold, \(\manifold\).
While the solution to~(\ref{eq:minimization}) is an equivalence class, we choose an arbitrary representative of the class since we are interested in the value of \(f\).
The closely related Stiefel manifold consists of the same problem without the homogeneity condition. While the Householder transformation is suitable for both the Grassman and Stiefel manifolds, the optimization method presented does not optimize with respect to the basis and is therefore suitable only for the Grassman manifold.

In the applications we have in mind the evaluation of \(f\) and \(df\) is expensive. For this reason we cannot employ a high quality line search to decide the step length, and the optimization method must be robust.
Furthermore, we will measure the number of evaluations of \(f\) and \(df\) necessary to obtain a solution. One iteration of the optimization procedure requires the two evaluations, once to evaluate the solution candidate and once to construct a quadratic approximation along the search direction.

For computational reasons we choose an \(m \times n\) matrix, \(\X\), as a representative of a point on \(\manifold\), and enforce the requirement 
\begin{equation}\label{eq:constraint}
\X^T\X = \id.
\end{equation}
While this representation includes more degrees of freedom than strictly necessary, the approach is suitable for use in practice~\cite{edelman1998}.

We use the inner product for matrices
\begin{equation}\label{eq:innerproduct}
(\A, \B ) = \mathrm{trace} (\A^T \B),
\end{equation}
and the tangent spaces at \(\X\) satisfying \(\X^T\X=\id\)
\begin{equation}
\{\Z=\X\A+\Y\, | \,\Y^T \X = \zero \;\mathrm{and}\; \A^T=-\A\}.
\end{equation}
On the Grassman manifold the value of \(f\) depends only on the space spanned by the columns of \(\X\). We can therefore ignore the \(\X\A\) component of the tangent space, and denote
\begin{equation}
\tangentspace{\X} = \{\Y\, | \,\Y^T \X = \zero \}.
\end{equation}
On the Grassman manifold it is possible to substitute equivalence classes for the representatives we have chosen, but practical computations require us to always use a specific matrix.

We also assume that we are given a direction \(\W\) by the minimization method in which we want to move on the manifold.
We project the direction on to \(\tangentspace{\X}\) by
\begin{equation}
\Y = (\id - \X\X^T)\W.
\end{equation}
The motivation for the problem under consideration comes from density functional theory (DFT) electronic structure calculations \cite{marx,saad}.
We construct a simultaneous Householder operator that can be used to ensure that the optimization method naturally enforces the orthogonality constraint.
This approach differs from several other approaches in that it does not solve the canonical electron orbitals \cite{kresse1,marzari,voorhis2002,cances2001,vandevondele2003,saad}, instead we only solve the electron density that would be given by the orbitals.
To obtain the canonical orbitals from the electron density a linear eigenvalue problem must then be solved in the space spanned by the columns of \(\X\).
A similar approach using polynomial filtering can be found in~\cite{zhou2006,bekas2008}.
It is also possible to solve a nonlinear eigenvalue problem instead of the minimization problem \cite{pulay,kresse1,luke,saad}.

In \cite{edelman1998} a framework for optimization methods on the Stiefel and Grassmann manifolds is presented, while~\cite{chu1983} discusses a Newton-like iteration scheme on a more general manifold.
Univariate optimization methods for the Stiefel manifold is presented in \cite{celledoni2008}, where identity plus rank one Householder transforms are given as one possible choice for moving on the manifold.
The choice of coordinates can also be based on a QR factorization and polar decompositions~\cite{celledoni2002,dieci2003} or Lie groups~\cite{krogstad2003}.
An overview of geometric numerical integration techniques can be found in~\cite{lubich_hairer_wanner}.

First, we present a simultaneous Householder transformation in Section~\ref{sec:householder} that we can use to move on both the Stiefel and Grassman manifolds.
Then in Section~\ref{sec:optimization} we recall the method of steepest descent, the quasi-Newton (QN), and the nonlinear conjugate gradient (NLCG) methods adapted for use with the Householder operator.
In Section~\ref{sec:numerics} we numerically demonstrate the method on a model problem that includes nonlinearities similar to a DFT problem.
Finally, Section~\ref{sec:conclusion} presents the conclusion.

\section{Householder operator}\label{sec:householder}
We ensure that \(\X\in\manifold\) during the solution process by using the Householder transformation to move from one solution candidate to the next.

\begin{figure}
\begin{center}
\includegraphics[width=0.9\textwidth]{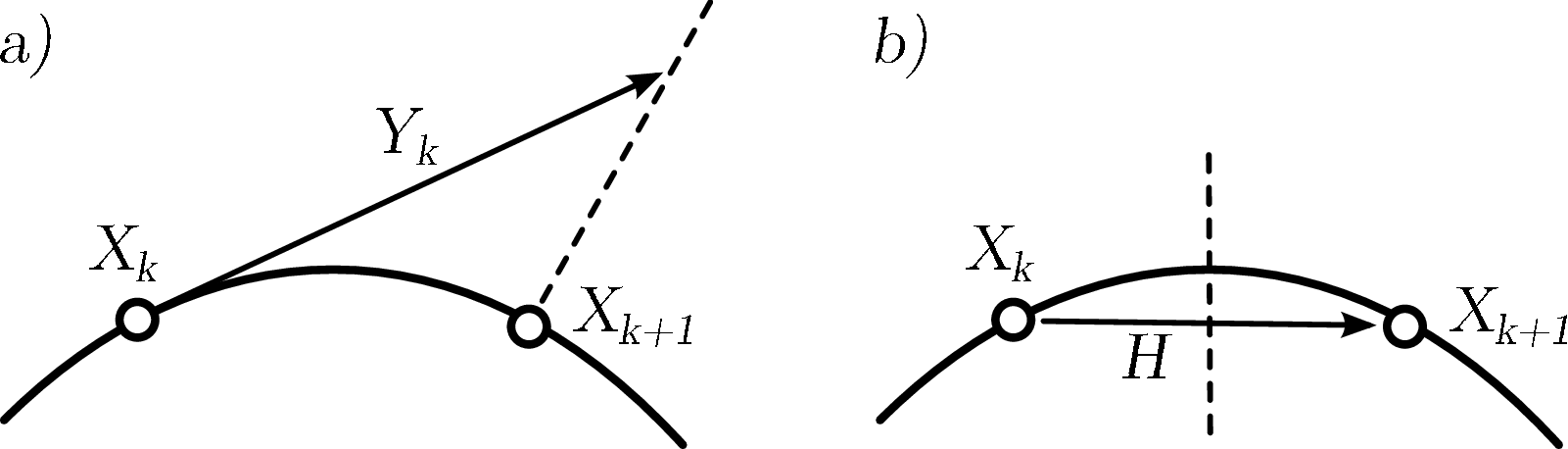}
\caption{\label{fig:update}Conceptual difference between reorthogonalization and Householder approach. In Subfigure a) the step is taken without regard to the manifold, and after the step is taken the new solution candidate \(\X_{k+1}\) is constructed by reorthogonalizing \(\X_k+\Y_k\). Subfigure b) illustrates the case where the Householder operator, \(\householder\), constructs an update \(\X_{k+1}\) that immediately satisfies the orthogonality condition.}
\end{center}
\end{figure}

To do this we need to find an operator
\begin{equation}\label{eq:householder}
\householder(\step) = \id - 2 \Q(\step)\Q(\step)^T,
\end{equation}
where
\begin{equation}\label{eq:q_orth}
\Q(\step)^T\Q(\step) = \id\quad\forall \step,
\end{equation}
and \(\step\) is a parametrization of \(\householder\) such that
\begin{equation}\label{eq:qinit}
\householder(0)\X = \X.
\end{equation}
This requirement leads to
\begin{equation}\label{eq:q_init}
\Q(0)^T\X=\zero.
\end{equation}

\(\householder\) is unitary, and if we let
\begin{equation}
\X_{k+1} = \householder(\step)\X_k,
\end{equation}
we obtain a sequence \(\{\X_k\}\) that satisfies
\begin{equation}
\X_{k+1}^T\X_{k+1} = \X_k^T\X_k \quad \forall\; k, \step.
\end{equation}

The orthonormality requirement~(\ref{eq:q_orth}) on \(\Q(\step)\) leads to the constraint
\begin{equation}\label{eq:diff}
\frac{\partial\Q}{\partial\step}^T\Q + \Q^T\frac{\partial\Q}{\partial\step} = \zero.
\end{equation}

We set the initial condition for \(\tfrac{\partial\Q}{\partial \step}\) by requiring that
\begin{equation}\label{eq:qdotinit}
\frac{\partial}{\partial \step} \left( \householder(\step)\X\right)\Big|_{\step=0} = \Y,
\end{equation}
where \(\Y\in\tangentspace{\X}\) is a projected direction given by the minimization method we choose to employ.
Differentiating with respect to \(\step\), and using property~(\ref{eq:q_init}), we obtain from~(\ref{eq:qdotinit})
\begin{equation}
-2  \Q(0)\frac{\partial\Q}{\partial\step}(0)^T \X = \Y.
\end{equation}

We set
\begin{equation}\label{eq:v}
\Q(0) = \V,
\end{equation}
where \(\V\R\) is the compact QR decomposition of \(\Y\), and choose
\begin{equation}
\frac{\partial\Q}{\partial\step}(0) = - \tfrac{1}{2} \X\R^T.
\end{equation}

A solution satisfying Equation~\eqref{eq:diff} and conditions~(\ref{eq:qinit}) and~(\ref{eq:qdotinit}) is
\begin{equation}\label{eq:qsolve}
\tilde{\Q}(\step) = \tilde{\Q}_0 \exp\Bigl(\step\begin{bmatrix}\zero&\frac{1}{2}\R\\ -\frac{1}{2}\R^T&\zero\end{bmatrix}\Bigr).
\end{equation}
Here \(\tilde{\Q}_0=\begin{bmatrix}\V & \X\end{bmatrix}\) and \(\Q(\step)\) corresponds to the first \(n\) columns of \(\tilde{\Q}(\step)\).
Given any \(\Q\) with orthonormal columns \(\householder\) constructed by~(\ref{eq:householder}) ensures that \(\X_{k+1}^T\X_{k+1} = \id\). For this reason we also consider a Householder operator constructed from a second order expansion of the matrix exponential that has subsequently been orthonormalized by the QR~method to ensure that the orthogonality constraint is satisfied.
This approach is similar to the orbital transformation but includes orthogonalization after every evaluation of the matrix exponential function~\cite{vandevondele2003}. To distinguish these from the basic algorithms we use AEQN and AENLCG to denote the approximate exponential versions.

\textbf{Remark:} If \(m < 2 n\) the requirement \(\Y^T\X = \zero\) restricts the number of columns in \(\Y\) to below \(n\). In this case the size of the first block in Equation~(\ref{eq:qsolve}) should be reduced accordingly. A similar modification must be made if \(\Y\) is not full column rank. For the Householder transformation to work we must still have \(m > n\).

%\textbf{Remark:} It is also possible to orthonormalize \(\Y\) by left multiplication with \(\M^{-1/2}\), where \(\M = \Y^T\Y\). In this case Equation~(\ref{eq:qsolve}) remains the same if we set \(\V = \Y\M^{-1/2}\) and \(\R = \M^{1/2}\).

\section{Descent methods with orthogonality constraints}\label{sec:optimization}

In this section we consider the method of steepest descent, a quasi-Newton method, and a nonlinear conjugate gradient method for minimization with orthogonality constraints. We also present the Householder operator for an \(\overlap\)-orthonormal basis, and combine this with the optimization methods.

\subsection{The method of steepest descent}\label{sec:sd}
The method of steepest descent for the Stiefel manifold is also known as the projected gradient method~\cite{chu1990}.
At each step we simply set
\begin{equation}\label{eq:direction}
\Y_k = -\sigma(\id - \X_k\X_k^T)\nabla f(\X_k).
\end{equation}
The parameter \(\sigma > 0\) is almost redundant for the method of steepest descent, but will become important for the quasi-Newton methods presented later.

To decide the step length we evaluate \(f(\householder(\step_k^e)\X_k)\), where \(\step_k^e\) is an estimate step length, and construct the quadratic approximation \(p(\step)\) of \(f(\householder(\step)\X_k)\).
We then solve \(\step_\mathrm{min} = \mathrm{argmin}\;p(\step)\) from the system
\begin{align}\label{eq:step}
p(0) &= f(\householder(0)\X),\\
p(\step_k^e) &= f(\householder(\step_k^e)\X),\\
p'(0) &= (\nabla f(\householder(0)\X), \Y).
\end{align}
This permits us to compute \(\householder(\beta\step_\mathrm{min})\) and evaluate \(f(\householder(\beta\step_\mathrm{min})\X)\) as well as \(\nabla f(\householder(\beta\step_\mathrm{min})\X)\), where \(\beta\) is an underrelaxation parameter. To ensure that we obtain a non-increasing iteration we choose the step length \(\step_k\) from the set \(\{0, \beta\step_\mathrm{min}, \step_k^e\}\), such that we obtain the lowest value of \(f\) evaluated so far.
If \(\step_{k} = 0\) we set \(\step_{k+1}^e = 0.25\times\step_k^e\) and otherwise we set \(\step_{k+1}^e = \mathrm{min} (|\step_\mathrm{min}|, 2\,\step_k^e)\).

\textbf{Remark:} It turns out that \(\step_k^e\) often is an acceptable choice for step length, and we believe that it is possible to construct a completely line search free minimization method with adaptive step length \cite{baarman2011b}.
However, it must be tuned for a real-world problem, and we will not explore this option here.

\subsection{Vector transport on the manifold}

\begin{figure}
\begin{center}
\includegraphics[width=0.9\textwidth]{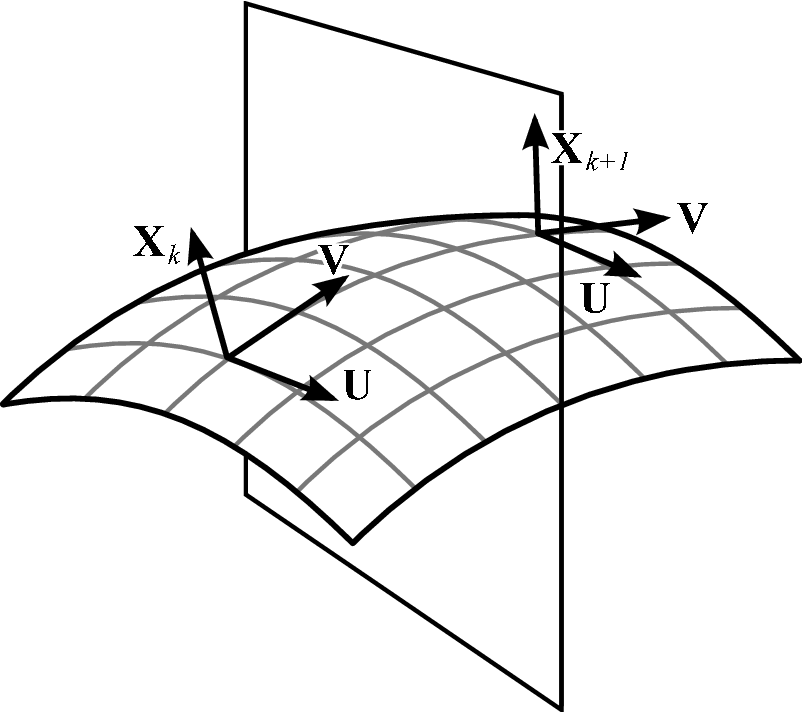}
\caption{\label{fig:transport}The curvature of the manifold must be taken into account when the trial solution is updated. The update operator, \(\householder\), corresponds to a reflection in the illustrated plane.}
\end{center}
\end{figure}

To reduce the number of iterations we use information from previous evaluations to improve the search direction.
To this end we construct a transport operator, \(\transport: \tangentspace{\X_k} \rightarrow \tangentspace{\X_{k+1}}\), that moves tangent vectors at \(\X_k\)to the tangent space at \(\X_{k+1}\).

Any given vector, \(\Z_k \in \Real^{m\times n}\), associated with a candidate solution, \(\X_k\), and search direction, \(\V\), can be decomposed into 
\begin{equation}\label{eq:decomposition}
\Z_k = \X_k\A + \V\B + \U\C,
\end{equation}
and we demand that \(\X_k^T\U = \V^T\U = \zero\), and \(\U^T\U = \id\) in addition to \(\X_k^T\V = \zero\) which is satisfied by construction. For simplicity we assume that \(\Z_k\) is such that \(\U\in\Real^{m\times n}\) and \(\A, \B, \C \in \Real^{n\times n}\).
\(\A\) and \(\B\) can be computed by projecting \(\Z_k\) onto \(\X_k\) and \(\V\) respectively, while \(\U\C\) is for example the QR-decomposition of the remainder.
In practice the decomposition is not explicitly constructed.

The different parts of the decompostion (\ref{eq:decomposition}) must be transported separately when the position is updated to \(\X_{k+1} = \householder \X_k\). The component spanned by \(\X_k\) is reflected correctly by \(\householder\), however a reflection gives the wrong sign to the \(\V\) component of \(\Z_k\). The final component \(\U\) is orthogonal to both \(\X_k\) and \(\V\), and should not change when \(\householder\) is applied. The transport operation is shown in Figure~\ref{fig:transport}.
We construct the transport operator by separating the components of \(\Z_k\) by projection and subsequent application of \(\householder\).
The transport operator therefore becomes
\begin{equation}
\householder\X_k\X_k^T - \householder\V\V^T + (\id - \X_k\X_k^T - \V\V^T).
\end{equation}
In practice we apply this for vectors \(\Z_{k} \in \tangentspace{\X_k}\) which satisfy \(\Z_{k}^T\X_k = \zero\), and we can use the simplified transport operator
\begin{equation}\label{eq:transport}
\transport(\step) = - \householder(\step)\V\V^T + (\id - \V\V^T),
\end{equation}
and the tangent vector corresponding to \(\Z_k\) at \(\X_{k+1}\) is
\begin{equation}
\Z_{k+1} = \transport(\step_k)\Z_{k} \in \tangentspace{\X_{k+1}}.
\end{equation}

\subsection{\(\overlap\)-orthonormal Householder and transport operator}
In practice \(\X\) is often represented by a discretization that requires a generalization of the orthonormality constraint~(\ref{eq:constraint}). The generalized constraint is
\begin{equation}\label{eq:sorth}
\X^T\overlap\X = \id,
\end{equation}
where \(\overlap\) is symmetric and positive definite.
This constraint arises for example as an overlap matrix in DFT or a mass matrix in finite element calculations.
If \(\overlap\) is sparse or has other structure that can be exploited it can be preferable not to change basis for the optimization procedure.
We therefore also present the Householder transform and optimization procedure for the \(\overlap\)-orthonormal case.
We can construct both the Householder transform and the transport operator using the same argument as for the regular orthonormal case, as long as we account for \(\overlap\) orthonormality~(\ref{eq:sorth}).
However, if \(\overlap\) is full, and lacks exploitable structure, the operation \(\overlap\X\) has asymptotic complexity \(m^2n\) and overshadows the rest of the procedure.

The projection of the direction of steepest descent onto the \(\overlap\)-orthogonal manifold is
\begin{equation}
\Y = -\sigma(\id - \X\X^T\overlap)\nabla f(\X)
\end{equation}
instead of~(\ref{eq:direction}) and satisfies \(\Y^T\overlap\X = \zero\). We must also use the \(\overlap\) weighted compact QR decomposition to compute a factorization \(\Y=\V\R\) where \(\V^T\overlap\V = \id\), and \(\R\) is upper triangular.

With these modifications, the Householder operator in~(\ref{eq:householder}) becomes
\begin{equation}
\householder_\overlap(\step) = \id - 2 \Q(\step)\Q(\step)^T\overlap,
\end{equation}
where \(\Q(\step)\) is as in Equation~(\ref{eq:qsolve}), and remains unchanged.
The \(\overlap\)-orthonormal transport operator corresponding to~(\ref{eq:transport}) is
\begin{equation}\label{eq:transport_operator}
\transport_\overlap(\step) = - \householder_\overlap(\step)\V\V^T\overlap + (\id -  \V\V^T\overlap).
\end{equation}

\subsection{The quasi-Newton method based on Householder transforms}
The method of steepest descent generally performs poorly if the minimum of the target function is at the bottom of a narrow valley.
Newton's method solves this problem, but requires that the Hessian of the function is available to determine the search direction.
When the Hessian is not available we can replace it with an approximation of the true inverse Hessian of the system to obtain a quasi-Newton method.

We base our method on Broyden's second or \emph{bad} update to construct the approximate inverse Hessian, \(\G_k\), of \(f\) at \(\X_k\). While Broyden's second update does not construct a symmetric approximation, or ensure that the approximation is positive definite it is a robust choice for electronic structure calculations \cite{luke,baarman2011a,baarman2011b}.
However, we must take into account that our vectors are actually \(\Real^{m\times n}\) matrices, which will lead to a method identical to the generalized Broyden update.
The secant condition is then
\begin{equation}\label{eq:secant}
\G_{k+1} \Delta \F_k = \Delta \X_k,
\end{equation}
where we project the orbital differences
\begin{equation}
\Delta\X_k = (\id - \X_{k+1}\X_{k+1}^T\overlap)(\X_{k+1}-\X_{k}),
\end{equation}
and gradient differences
\begin{equation}
\Delta\F_k = (\id - \X_{k+1}\X_{k+1}^T\overlap)\nabla f(\X_{k+1})-\transport(\step_k)(\id-\X_k\X_k^T)\nabla f(\X_{k})
\end{equation}
onto \(\tangentspace{\X_{k+1}}\).
The no change condition is now
\begin{equation}
\G_k\Z = \G_{k+1}\Z\quad \forall\,\Z\;:\,\Z^T\Delta \F_k = \zero.
\end{equation}
These conditions corresponds to the generalized Broyden's second update for groups of size~\(n\).
We can therefore use the generalized update formula \cite{fang}
\begin{equation}
\G_{k+1} = \G_{k} + (\Delta \X_k - \G_k \Delta \F_k) (\Delta \F_k^T\overlap\Delta\F_k)^{-1}\Delta\F_k^T\overlap.
\end{equation}
As initial guess we use \(\G_0 = \sigma \id\).
With this choice, the quasi-Newton method is identical with the method of steepest descent if we do not enforce any secant conditions~(\ref{eq:secant}). When secant conditions are enforced we can use \(\sigma\) to control the influence of \(\G_0\) compared to the information gained from the secant conditions. In general, the information gained from these is reliable, and therefore \(\sigma\) should be small \cite{baarman2011a,baarman2011b}.

To construct the search direction we use
\begin{equation}
\Y_k = -\G_k(\id - \X_k\X_k^T\overlap)\nabla f(\X_k).
\end{equation}

In practice, we do not store \(\G_k\) as a full matrix. Instead we represent it as a low rank update.
Details on recursive or low rank implementation of \(\G_k\) can be found in \cite{kawata,nocedal,baarman2011a,baarman2011b}.

We also limit the number of secant conditions used to construct \(\G_k\).
Each condition requires storage of two \(m\times n\) matrices, \(\Delta \X\) and \(\Delta \F\), and these matrices must be transported to \(\tangentspace{\X_k}\) after each step.
This is done with the transport operator, \(\transport_\overlap(\step)\), defined in Equation~(\ref{eq:transport_operator}).
As we demonstrate in Section~\ref{sec:numerics} the first few secant conditions offer dramatic improvement over the method of steepest descent, but further secant conditions do not give the same benefit.
For this reason we limit the secant conditions by a pre-determined history length, and simply discard older conditions.

We use the same line search as the one presented in Section \ref{sec:sd}.
If, however
\begin{equation}
(\Y_k,\overlap(\id - \X_k\X_k^T\overlap)\nabla f(\X_k)) \ge 0
\end{equation}
then the proposed direction is not a descent direction. In this case we restart the optimization method and forget the secant history.
If the line search returns the current point \(\X_k\), we update \(\G_k\) but stay at \(\X_k\).

\subsection{Nonlinear conjugate gradients}\label{sec:nlcg}
The linear conjugate gradient (CG) method can be viewed as a optimization method for a quadratic problem. Several generalizations of the CG method have been presented to solve optimization problems that are not of quadratic form~\cite{nocedal}.
Below, we review a nonlinear CG method adapted to account for the curvature of the manifold~\cite{edelman1998}.

Given \(\X_0\) which satisfies \(\X_0^T\X_0 = \id\), the gradient projected onto \(\tangentspace{\X_0}\) is
\begin{equation}
\Y_0 = (\id - \X_0\X_0^T\overlap)\nabla f(\X_0),
\end{equation}
and the initial search direction is the direction of steepest descent
\begin{equation}
\direction_0 = -\Y_0.
\end{equation}

On the manifold the NLCG method then proceeds by minimizing \(f\) along the path defined by the search direction \(\direction_k\). In practice we evaluate \(f\) once along the search direction and minimize the quadratic approximation as in Section~\ref{sec:sd}.
The next candidate is chosen as the best evaluated step, \(\step_k\),
\begin{equation}
\X_{k+1} = \householder(\step_k)\X_k,
\end{equation}
and the gradient and conjugate directions are transported to \(\tangentspace{\X_{k+1}}\) by \(\transport(\step_k)\) in Equation~\eqref{eq:transport_operator}.
The new gradient
\begin{equation}
\Y_{k+1} = (\id - \X_{k+1}\X^T_{k+1}\overlap)\nabla f(\X_{k+1}),
\end{equation}
and conjugate direction
\begin{equation}
\direction_{k+1} = -\Y_{k+1} + \gamma_k \transport_\overlap (\step_k) \direction_k,
\end{equation}
are then computed where
\begin{equation}
\gamma_k = \frac{(\Y_{k+1}-\transport_\overlap(\step_k)\Y_k,\Y_{k+1})}{(\Y_k,\Y_k)}.
\end{equation}

For comparison we also implement a projected NLCG (PNLCG) method. Instead of ensuring that orbital updates satisfy \(\X_{k+1}^T\X_{k+1} = \id\) we orthogonalize \(\X_{k+1}\) after every update with the QR method. The conjugate directions and gradient are not transported to \(\tangentspace{\X_{k+1}}\), instead they are updated with the \(\id - \X_{k+1}\X_{k+1}^T\overlap\) projector onto \(\tangentspace{\X_{k+1}}\).

\section{Numerical experiments}\label{sec:numerics}
We use a two dimensional model problem with the condition \(\X^T\overlap\X = \id\) to compare the projected NLCG method with the NLCG and QN methods where satisfaction the orthonormality condition is ensured by the update operator.
The model problem is inspired by electronic structure theory, and corresponds to a three dimensional system constrained to two dimensions without exchange-correlation terms.

The target function is \cite{marx}
\begin{equation}\label{eq:model_problem}
f(\X) = -\tfrac{1}{2}\mathrm{tr}((\overlap^{1/2}\X)^T\laplace\overlap^{1/2}\X) + \extpotential^T\density+\tfrac{1}{2} \density^T\intpotential\density,
\end{equation}
where \(\laplace\in\Real^{m\times m}\) is the discretized Laplace operator, \(\extpotential\in\Real^{m}\) the external potential, \(\density \in\Real^m\) the electron density, and  \(\intpotential\density\) the Hartree potential. The electron density is
\begin{equation}
\density_i = \sum_{j=1}^n((\overlap^{1/2}\X)\circ(\overlap^{1/2}\X))_{ij},
\end{equation}
where \(\circ\) is the entrywise, or Hadamard, product.
We use the overlap matrix
\begin{equation}
\overlap = \frac{1}{9h^2}
\begin{bmatrix}
\M & \tfrac{\M}{4} & 0 & \cdots\\
\tfrac{\M}{4} & \M & \tfrac{\M}{4} & \ddots \\
0 & \tfrac{\M}{4} & \M & \ddots \\
\vdots & \ddots & \ddots & \ddots \\
\end{bmatrix},
\end{equation}
where \(h\) is the one dimensional grid size and
\begin{equation}
\M = 
\begin{bmatrix}
4 & 1 & 0 & \cdots\\
1 & 4 & 1 & \ddots \\
0 & 1 & 4 & \ddots \\
\vdots & \ddots & \ddots & \ddots \\
\end{bmatrix}.
\end{equation}
This corresponds to the mass matrix of a finite element discretization with bilinear quadratic element. The calculations have also been performed with a symmetric and positive definite random matrix.
%We use a random symmetric and positive definite overlap matrix, \(\overlap\).
It turns out that as long as \(\overlap\) is well conditioned, it has only a small effect on the rate of convergence.

To calculate the potentials we use
\begin{equation}
\extpotential_i = -\sum_{j=1}^N \frac{Z_j}{||\position_i - \Position_j||+\alpha},
\end{equation}
where the sum is over the nuclei with charge \(Z_j\) and position \(\Position_j\). The position corresponding to the discretization point \(i\) is \(\position_i\), and the parameter \(\alpha\) is used to regularize the potential. \(\intpotential\in\Real^{m\times m}\) is similarly given by
\begin{equation}
\intpotential_{ij} = \frac{1}{||\position_i-\position_j|| + \alpha}.
\end{equation}

We solve the problem in the unit square with zero boundary conditions corresponding to an infinite potential well. We use a uniform finite difference discretization with \(m\) inner points to obtain a system where \(\X \in \Real^{m\times n}\). Here \(n\) corresponds to the number of electrons.
As initial guess we use the solution of the quadratic problem using the first two terms of~(\ref{eq:model_problem}), and choose \(\step_0^e = 1.0\) to initialize the minimization procedure, cf.~Equation~(\ref{eq:step}).

As generators for the external potential we use two nuclei, where one is placed at the grid point closest to \((\tfrac{1}{3},\tfrac{1}{3})\), and the other at the grid point closest to \((\tfrac{2}{3},\tfrac{13}{24})\). The off diagonal placement is chosen to break the symmetry of the system.
We use
\begin{equation}
\epsilon_\X = ||(\id - \X\X^T)\nabla f(\X)||/\sqrt{mn},
\end{equation}
to measure convergence, and consider the system converged when
\begin{equation}
\epsilon_\X < 10^{-2}.
\end{equation}
At this point
\begin{equation}
|f(\X_\mathrm{Ref})-f(\X)| \approx 10^{-4},
\end{equation}
where the reference solution has been calculated such that
\begin{equation}
\epsilon_{\X_\mathrm{Ref}} < 10^{-5}.
\end{equation}

%\begin{figure}
%\begin{center}
%\includegraphics{Fig3}
%\caption{\label{fig:colorpic}Electron density of a 6 electron two dimensional system with two nuclei. The nuclei at \((\tfrac{1}{3},\tfrac{1}{3})\) has a charge \(Z_1 = 3\), and the charge of the nuclei at \((\tfrac{2}{3},\tfrac{13}{24})\) is \(Z_2 = 3\). The isocurves correspond to the black cutoffs in the legend. The parameters used for the calculation are \(m= 40^2\), \(n=6\), \(\beta=0.4\), \(\sigma = 10^{-4}\), \(\alpha = 2\times10^{-2}\), and history length 6. Convergence is reached in 36 iterations with the quasi-Newton method.}
%\end{center}
%\end{figure}
%A solution to Equation~(\ref{eq:model_problem}) with \(n=6\) is presented in Figure~\ref{fig:colorpic}. This solution exhibits several properties that can also be found in the full electronic structure problem.
%The core electrons are localized around the nuclei, and the valence orbitals are spread out.
%and at three other positions further from the nuclei.
%Furthermore, the nucleus with greater charge has a lower peak in electron density than the less charged nucleus.
%This is due to higher order terms in~(\ref{eq:model_problem}), which result in partial screening of the larger nucleus.

\begin{figure}
\begin{center}
\subfigure{\includegraphics[width=0.45\textwidth]{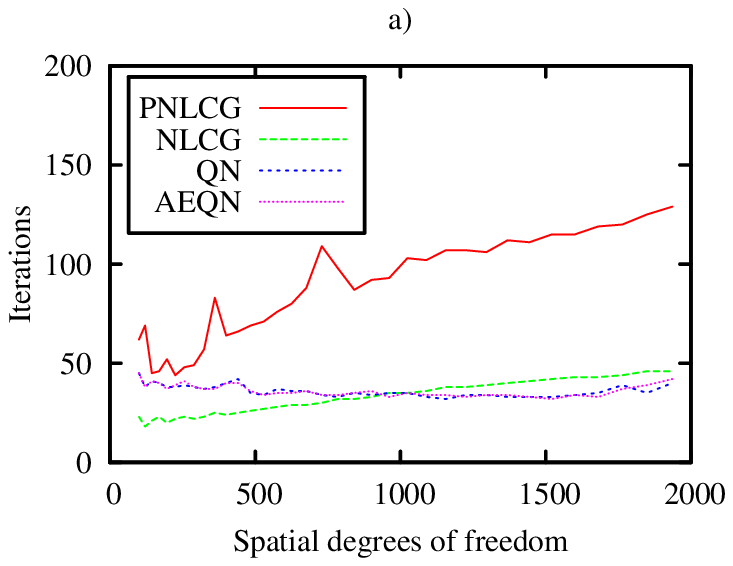}}
\subfigure{\includegraphics[width=0.45\textwidth]{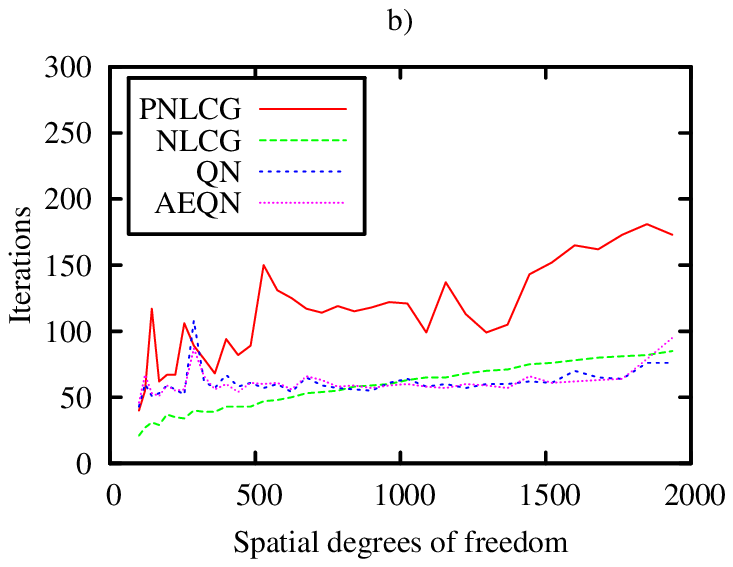}}
\caption{\label{fig:dof}Iterations required for convergence as a function of the spatial degrees of freedom, \(m\). For subfigure a) the external potential is generated by two nuclei, \(Z_1 = 3\), \(Z_2 = 3\), and the parameters of the calculation are \(n=6\), \(\beta = 0.5\), \(\sigma = 10^{-4}\), \(\alpha = 2\times 10^{-2}\), and history length is 6. For subfigure b) \(Z_1 = 4\), \(Z_2 = 3\), and \(n = 7\).
Here NLCG corresponds to the nonlinear conjugate gradient method, PNLCG to the projected NLCG, QN to the quasi-Newton method and AEQN to the approximate exponent QN method.
The number of iterations required for convergence is identical for the AENLCG and NLCG methods, where AENLCG is the approximate exponent NLCG method.}
\end{center}
\end{figure}
Figure \ref{fig:dof} presents the iterations necessary for convergence for a six and seven electron system.
These iterations roughly grows as the square root of the degrees of freedom for the NLCG methods.
However, the projected NLCG method performs significantly worse than the NLCG and QN methods adapted for the manifold.
It turns out that the AENLCG method requires the same number of iterations to converge as the NLCG method, while a small difference is visible between QN and AEQN methods.

\begin{figure}
\begin{center}
\subfigure{\includegraphics[width=0.45\textwidth]{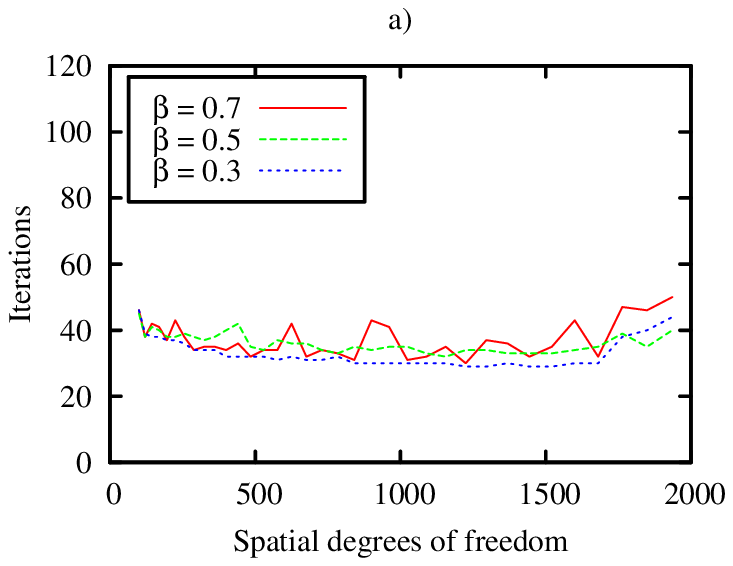}}
\subfigure{\includegraphics[width=0.45\textwidth]{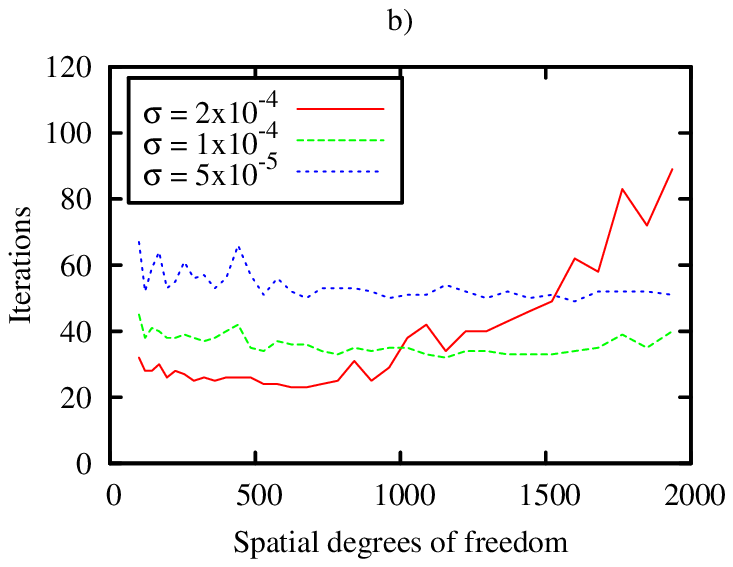}}
\caption{\label{fig:parameters}Iterations required for convergence of the QN method as a function of the spatial degrees of freedom, \(m\). For the both figures the external potential is generated by two nuclei, \(Z_1 = 3\), \(Z_2 = 3\), and the parameters of the calculation are \(n=6\), \(\alpha = 2\times 10^{-2}\), and history length is 6.
Unless otherwise indicated in the figure \(\beta = 0.5\) and \(\sigma = 10^{-4}\).
}
\end{center}
\end{figure}

\begin{figure}
\begin{center}
\subfigure{\includegraphics[width=0.45\textwidth]{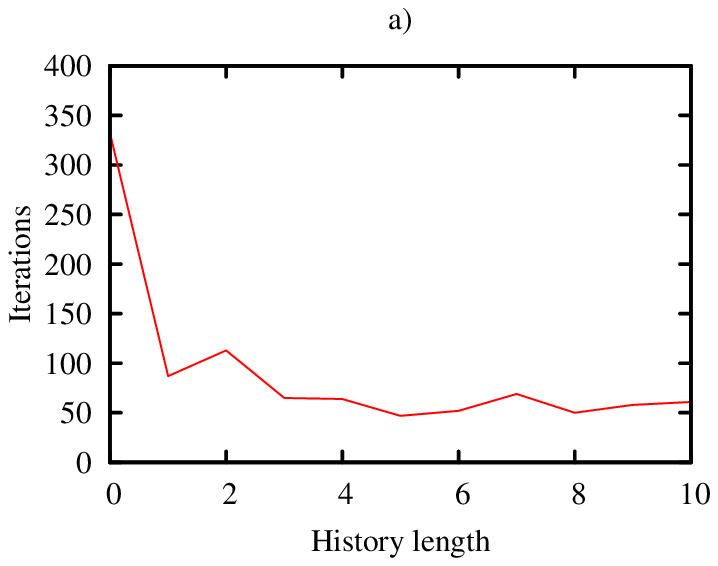}}
\subfigure{\includegraphics[width=0.45\textwidth]{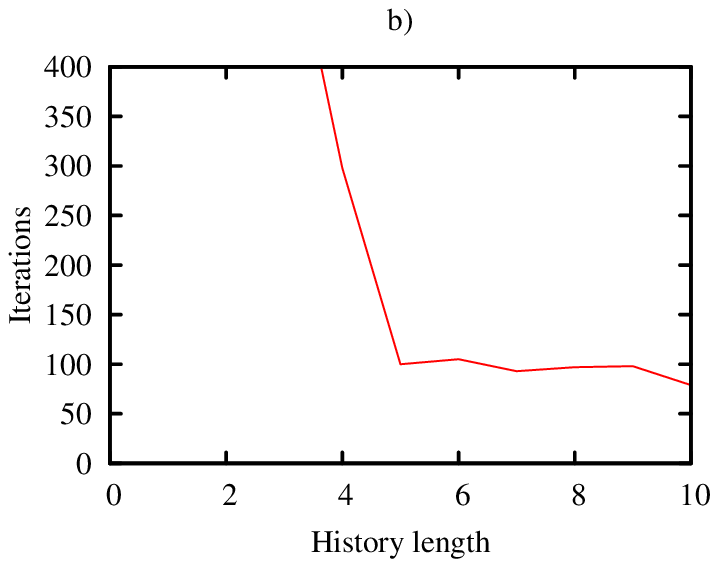}}
\caption{\label{fig:history}Iterations required for convergence as a function of history length for the QN method for a six and seven electron system for subfigure a) and b) respectively. Spatial degrees of freedom is 2500 and history length varies, parameters are otherwise identical to Figure~\ref{fig:dof}.
%The external potential is generated by two nuclei, \(Z_1 = 3\), \(Z_2 = 3\), and the parameters of the calculation are \(m=35^2\), \(n=7\), \(\beta=0.5\), and \(\sigma = 5\times10^{-5}\).
%The method fails to converge in 500 iterations for history length shorter than 2.}
}
\end{center}
\end{figure}

In Figure~\ref{fig:parameters} the effect of weight, \(\sigma\), of the initial approximation of the inverse Hessian and the underrelaxation, \(\beta\) are presented.
From Figure~\ref{fig:parameters}~b) it is clear that \(\sigma\) is particularly important for fast convergence of the quasi-Newton method.
The effect of \(\beta\) is much smaller, and while \(\beta\) can in some cases improve convergence the effect of \(\sigma\) is significantly more important.
A low \(\sigma\) results in slower convergence as long as the more aggressive parameter choice converges well.
However, when the rate of convergence begins to suffer from the more aggressive parameter choice the rate of convergence can be improved by a more conservative choice.
These results agree with earlier work \cite{baarman2011a}, which indicate that the secant conditions offer reliable information of the electronic structure problem, while the initial approximation of \(\G\) is less reliable.
The history length of the QN method must also be sufficient for the method to perform well. This is illustrated in Figure~\ref{fig:history}.

\section{Conclusion}\label{sec:conclusion}

We have presented a Householder update scheme which ensures that the columns remains orthogonal, and is suitable for both NLCG and QN methods.
Furthermore, the operator allows us to transport gradient information and construct secant conditions from previous evaluations of \(f\) to the tangent space of the best candidate solution.
This approach eliminates the need to parametrize the manifold, and permits us to use standard linear algebra routines to update the solution candidate.

We have demonstrated the methods numerically on a model problem inspired by the electronic structure problem, and compared them to a projected NLCG method.
Taking the underlying manifold into account significantly improves convergence rate of the optimization methods, and using a second order orthonormal approximate matrix exponent does not decrease performance of the QN or NLCG methods.

The QN method is significantly improved by taking the first few secant conditions into account when constructing the approximation of the Hessian of \(f\).
However, for the secant condition history to improve convergence speed of the QN method the manifold must be taken into account.
The update of the secant conditions is also based on the Householder operator that is used to update the solution candidate.
While the performance of the QN method depends on the weight of the initial approximate Hessian the QN method performs well once the weight is correctly set.

While the QN method is sensitive to the correct choice of \(\sigma\), the performance of the NLCG method does not depend on parameter choice.
Furthermore, the NLCG method is relatively simple to implement, and only requires a one step history. For these reasons we believe that the NLCG method is a good general purpose optimization method for electronic structure problems if the method is adapted to the manifold.

\section{Acknowledgments}
We are grateful towards Dr.~Mika~Juntunen for suggestions and comments on the manuscript.

\bibliographystyle{plain}
\bibliography{references}

\end{document}